\documentclass[aps,prd,twocolumn,groupedaddress,showpacs,nofootinbib,amssymb]{revtex4-2}
\usepackage[dvips]{graphicx}
\usepackage{amssymb}
\usepackage{amsmath}
\usepackage{graphicx,,color}
\usepackage{amsfonts}
\usepackage{bm}
\usepackage{cancel}
\usepackage{comment}

\newcommand\be{\begin{equation}}
\newcommand\ee{\end{equation}}

\newcommand\e{\mathrm{e}}

\allowdisplaybreaks[4]

\begin{document}

\tolerance=5000

\title{Kinetic Coupling Corrected Einstein-Gauss-Bonnet Gravity Late-Time Phenomenology}
\author{F.P.
Fronimos\,\thanks{fotisfronimos@gmail.com}}
\affiliation{ Department of Physics, Aristotle University of
Thessaloniki, Thessaloniki 54124,
Greece}

\tolerance=5000

\begin{abstract}
In this short note we present the dynamics of a general scalar-tensor model, and in particular a scalar Einstein-Gauss-Bonnet model with a non-minimal coupling between gravity and the kinetic term of the scalar field. For the sake of simplicity two $f(R)$ models are studied separately, an exponential and a power-law, accompanied by either an exponential or quartic scalar potential and a strictly exponential Gauss-Bonnet scalar coupling function known for being a suitable candidate for describing both the early and the late time. By introducing the general framework of a late-time study for an arbitrary scalar-tensor model, we find that the aforementioned models are capable of producing compatible with the Planck data observations and are in a relatively good agreement with the $\Lambda$CDM model and the GW170817 event as the tensor perturbation velocity is equal to unity in natural units for the whole are of values of redshift studied if certain parameters are properly designated. A brief comment on the appearance of dark energy oscillations which appear for the case of power-law $f(R)$ and the overall viability of the model is also made. 
\end{abstract}


\maketitle

\section{Introduction}
Currently one of the greatest mysteries that theoretical cosmologists strive to explain is the apparent accelerating expansion of the universe \cite{Riess:1998cb}. The formalism of general relativity however is incapable of describing such phenomenon without the inclusion of a cosmological constant. For this purpose, a variety of well motivated theories of modified gravity have recently emerged in order to account for such acceleration, for a detailed review see \cite{Nojiri:2019fft,Delubac:2014aqe,Sahni:2014ooa,Nojiri:2017ncd,Nojiri:2009kx,Capozziello:2011et,Capozziello:2010zz,Nojiri:2006ri,Nojiri:2010wj,delaCruzDombriz:2012xy,Olmo:2011uz,Nojiri:2003ft,Nojiri:2007as,Nojiri:2007cq,Cognola:2007zu,Nojiri:2006gh,Appleby:2007vb,Zhong:2018tqn,Odintsov:2020qyw,Li:2007jm,Nojiri:2005jg,Nojiri:2005am,Cognola:2006eg,Elizalde:2010jx,Izumi:2014loa,Oikonomou:2016rrv,Kleidis:2017ftt,Escofet:2015gpa,Makarenko:2017vuk,Makarenko:2016jsy,Navo:2020eqt,Bajardi:2020osh,Capozziello:2019wfi,Benetti:2018zhv,Clifton:2006kc,Bogdanos:2009tn,Capozziello:2004us,Barrow:1988xh,Bamba:2009uf,DeLaurentis:2015fea,DeFelice:2010sh,delaCruzDombriz:2011wn,Elizalde:2010ts,Odintsov:2020nwm,Bamba:2012qi}. Such theories may admit higher order corrections of gravity from curvature invariants or assume the existence of a scalar field which may be minimally or non minimally coupled to the aforementioned curvature invariants. The latter category is the so called scalar-tensor theory frequently seen in the literature and poses a plausible scenario since it manages to describe both the early and the late time era.

An interesting class of scalar-tensor theories which is also the main focus of the current article is the string inspired gravity. By admitting the existence of a scalar field such theories express low-energy quantum corrections in the gravitational action which in turn appear in the equations of motion non trivially, for a detailed review see \cite{Odintsov:2020sqy,Oikonomou:2020sij,Oikonomou:2020tct,Oikonomou:2020tct,Odintsov:2020ilr,Venikoudis:2021oee,Venikoudis:2021irr}. Here we shall focus mainly on a subclass of Horndeski theories which admits non minimal couplings between to two curvature terms. Firstly we make use of a nontrivial coupling between the scalar field and the Gauss-Bonnet topological invariant $\mathcal{G}$ which itself serves as higher order curvature correction. The non minimal coupling is essential given that the Gauss-Bonnet topological invariant is a total derivative in a 4D space. The second non minimal coupling refers to the kinetic term and the scalar field itself to the Einstein tensor. For simplicity the auxiliary scalar function which participates in non minimal kinetic coupling shall be the same as the Gauss-Bonnet scalar coupling function for simplicity. Such model as expressed before admits low-energy quantum corrections hence it is intriguing to examine the impact of such terms in the late-time era and examine whether their existence suffices in order to properly describe an accelerating expansion similar to what is currently experienced. The aforementioned curvature corrections are chosen since these terms are known for affecting the propagation velocity of gravitational waves hence it is interesting to examine whether a compatible with the $\Lambda$CDM model may actually be at variance with the GW170817 event or even violate causality due to such corrections. In the present article two such models shall be examined and the gravitational wave velocity shall be derived as a function of redshift in order to ascertain whether compatibility can be achieved. A thorough analysis is available in subsequent sections.

The main premise is to follow the steps as in Ref\cite{Odintsov:2020vjb,Odintsov:2021nim} where for a general modified theory of gravity, dark energy obtains a fundamental definition. Instead of being an arbitrary perfect fluid now dark energy serves as the effect of all geometric terms participating initially in the gravitational action that rectify Einstein's description. To showcase this , one performs two changes in the Friedmann equation and the continuity equation of the scalar field, which shall be numerically solved. The first change refers to a variable change and instead of cosmic time $t$, redhsift shall be used since it is more flexible. Afterwards all functions and their variations are rewritten appropriately using redhsift by using a differential operator that connects time with redshift, assuming that Hubble's parameter is a function of redshift. The second change involves Hubble and instead of using it directly, a new dimensionless statefinder $y_H$ shall be used which is connected to the dark energy density and facilitates the numerical solution of the aforementioned equations. Once the solutions for $y_H$ and the scalar field $\phi$ as functions of redshift are derived, a variety of statefinders such as the EoS of dark energy and the deceleration parameter are examined in order to ascertain whether the model at hand is compatible with observations. In general, both dark energy density and pressure are specified directly from the Friedmann-Raychaudhuri equations and in consequence both the equation of state and the density parameter are directly derivable from such terms. The model at hand shall be studied and essentially compared with the $\Lambda$CDM model, currently the most successful cosmological model that manages to describe observables to a really good extend.

The paper is constructed as follows: In section II the essential features of a late-time description for a general scalar-tensor model are discussed. By introducing certain auxiliary variables and functions that simplify to an extend the solution of the equations of motion, namely redshift $z$ and parameter $y_H$ as a replacement of cosmic time $t$ and Hubble's parameter respectively, various statefinder parameters are specified in order to study the phenomenological implications of the scalar-tensor model augmented by string corrective terms. Afterwards in section III the specific model of interest is specified and the equations of motion are solved numerically. We limit our work to a simple exponential and power-law $f(R)$ gravity separately in the presence of an exponential or quartic scalar potential respectively and an exponential Gauss-Bonnet scalar coupling function as such choices are known for describing the early era decently. Subsequently results are extracted for each and every statefinder and a comparison with the $\Lambda$CDM model  is performed \cite{Aghanim:2018eyx}. In addition the propagation velocity of gravitational waves is also examined in both models and a direct comparison with the GW170817 event is performed for the sake of consistency. Finally in section IV discussions are followed in which the results extracted are interpreted and compared with other late-time studies of similar nature.

\section{Aspects of Late-Time dynamics for a general scalar-tensor framework}
In the present paper we shall focus on the late-time study of a string inspired Einstein-Gauss-Bonnet theory or in other words a string inspired scalar-tensor theory. For this particular model, the gravitational action takes the following form

\begin{widetext}
\begin{equation}
\centering
\label{action}
S=\int{d^4x\sqrt{-g}\left(\frac{f(R,\phi)}{2\kappa^2}-\frac{1}{2}\nabla^\mu\phi\nabla_\mu\phi-V(\phi)-\xi(\phi)\mathcal{G}-\xi(\phi)cG^{\mu\nu}\nabla_\mu\phi\nabla_\nu\phi+\mathcal{L}_{matter}\right)}\, ,
\end{equation}
\end{widetext}
where $R$ denotes tha Ricci scalar and $f(R,\phi)$ is an arbitrary function for the time being depending solely on $R$ and the scalar field $\phi$, $g$ is the determinant of the metric tensor $g^{\mu\nu}$, $\kappa=\frac{1}{M_P}$ is the gravitational constant with $M_P$ being the reduced Planck mass, $\frac{1}{2}g^{\mu\nu}\nabla_\mu\phi\nabla_\nu\phi$ and $V(\phi)$ are the kinetic term and the corresponding potential of the scalar field $\phi$, $\xi(\phi)$ signifies the Gauss-Bonnet scalar coupling function since it is coupled to the Gauss-Bonnet topological invariant $\mathcal{G}$ defined as $\mathcal{G}=R_{\mu\nu\sigma\rho}R^{\mu\nu\sigma\rho}-4R_{\mu\nu}R^{\mu\nu}+R^2$ with $R_{\mu\nu\sigma\rho}$ and $R_{\mu\nu}$ being the Riemann curvature and Ricci tensor respectively, $c$ serves as an auxiliary constant introduced mainly for dimensional purposes such that it has mass dimensions $[m]^{-2}$ and finally $G^{\mu\nu}$ specifies the Einstein tensor given by the expression $G^{\mu\nu}=R^{\mu\nu}-\frac{1}{2}g^{\mu\nu}R$. Note that the $f(R,\phi)$ function is not specified here since the subsequent steps are general and do not require a particular designation to be valid however afterwards certain models of interest shall be chosen. Furthermore, for the scalar field in particular we shall assume that it is homogeneous, therefore $\phi=\phi(t)$ and given that the cosmological background in the case at hand is that of a flat Friedmann-Robertson-Walker (FRW) metric with the corresponding the line element being

\begin{equation}
\centering
\label{line}
ds^2=-dt^2+a^2(t)\delta_{ij}dx^idx^j\, ,
\end{equation}
where $a(t)$ denotes the scale factor, the kinetic term in consequence is simplified as $-\frac{1}{2}\dot\phi^2$ with the dot as usual implying differentiation with respect to cosmic time $t$. Moreover, due to the flat metric, the curvature invariants $R$ and $\mathcal{G}$ are written in terms of Hubble's parameter $H=\frac{\dot a}{a}$ as $R=6(2H^2+\dot H)$ and $\mathcal{G}=24H^2(\dot H+H^2)$ respectively. Thus, the equations of motion are simplified to an extend. Specifically, by implementing the variation principle with respect to the metric tensor $g^{\mu\nu}$ and the scalar field $\phi$, the field equations for gravity and the continuity equation are obtained which in this case have the following form

\begin{widetext}
\begin{equation}
\centering
\label{motion1}
\frac{3f_RH^2}{\kappa^2}=\rho_m+\frac{1}{2}\dot\phi^2+V+\frac{Rf_R-f}{2\kappa^2}-\frac{3H\dot f_R}{\kappa^2}+24\dot\xi H^3-9c\xi H^2\dot\phi^2\, ,
\end{equation}

\begin{equation}
\centering
\label{motion2}
-\frac{2f_R\dot H}{\kappa^2}=\rho_m+P_m+\dot\phi^2+\frac{\ddot f_R-H\dot f_R}{\kappa^2}-8H^2(\ddot\xi-H\dot\xi)-16\dot\xi H\dot H+2c\dot\phi\left(\xi(\dot H-3H^2)\dot\phi+2H\xi\ddot\phi+H\dot\xi H\dot\phi\right)\, ,
\end{equation}

\begin{equation}
\centering
\label{motion3}
\ddot\phi+3H\dot\phi+V_\phi-\frac{f_\phi}{2\kappa^2}+\xi_\phi\mathcal{G}-3c\left(H^2(\dot\xi\dot H+2\xi\ddot\phi)+2H(2\dot H+3H^2)\xi\dot\phi\right)=0\, ,
\end{equation}
\end{widetext}
where subscript $X$ implies partial differentiation with respect to the object $X$, here taking the values $R$ and $\phi$. These are the general equations of motion for the case of a non minimally coupled $f(R,\phi)$ gravity in the presence of a scalar potential and extra string corrections. The only string terms that are neglected are extra contribution from a higher order kinetic term. The equations of motion may seem lengthy but as mentioned are the general expressions. In the following, we shall study two cases of simple model functions as well where $f(R,\phi)=Re^{\frac{R_0}{R}}$ with $V(\phi)=\Lambda^2(1-e^{\frac{\phi}{\phi_0}})$ as a first example and $f(R,\phi)=\left(1+\left(\frac{\phi}{100M_P}\right)^2\right)R-2\Lambda\left(\frac{R}{\Lambda}\right)^{\frac{1}{100}}$ with $V(\phi)=(10\Lambda)^2\left(\frac{10^{38}\phi}{M_P}\right)^4$ afterwards with the Gauss-Bonnet scalar coupling function in both cases being described by an exponential function $\xi(\phi)=\e^{-\frac{\phi}{\phi_0}}$ for the sake of simplicity. All auxiliary parameters will be designated shortly however for the time being they are left unspecified. Let us proceed now with the introduction of redshift and certain auxiliary statefinder functions which facilitate our study.

It is convenient to perform a variable change and essentially discard cosmic time $t$ and in its place introduce redshift. In common literature, we define redshift as

\begin{equation}
\centering
\label{z}
1+z=\frac{1}{a(t)}\, ,
\end{equation}
hence time dependence of each function which appeared previously is connected with redshift. Note that this notation implies that the current scale factor is actually equal to unity. From this particular relation one can introduce a new differential operator as

\begin{equation}
\centering
\label{dz}
\frac{d}{dt}=-H(1+z)\frac{d}{dz}\, ,
\end{equation}
where hereafter it is assumed that $H=H(z)$. As a result, certain objects in the equations of motion (\ref{motion1}) through (\ref{motion3}) are rewritten as 

\begin{equation}
\centering
\label{dotH}
\dot H=-H(1+z)H'\, ,
\end{equation}

\begin{equation}
\centering
\label{dotfR}
\dot f_R=-H(1+z)f_R'\, ,
\end{equation}

\begin{equation}
\centering
\label{ddotfR}
\ddot f_R=H^2(1+z)^2f_R''+HH'(1+z)^2f_R'+H^2(1+z)f_R'\, ,
\end{equation}

\begin{equation}
\centering
\label{dotR}
\dot R=6H(1+z)^2\left(H'^2+H''-\frac{3HH'}{1+z}\right)\, ,
\end{equation}

\begin{equation}
\centering
\label{dotphi}
\dot\phi=-H(1+z)\phi'\, ,
\end{equation}
where prime implies differentiation with respect to redshift $z$ for simplicity. This transformation is universal for every object as it can easily be inferred. For the time derivative of $f_R$, one can also follow a different approach. Since $f=f(R,\phi)$ in general and $R=R(t)$, $\phi=\phi(t)$, we have

\begin{equation}
\centering
\label{dotf}
\dot f_X=\sum_{Y}f_{XY}\dot Y\, ,
\end{equation}
where $X,Y$ range from $R$ to $\phi$. Thus time dependence is now altered to $z$ dependence for every object. Therefore equations (\ref{motion1}) and (\ref{motion3}) can be solved as a system of second order differential equations with respect to Hubble's parameter and the scalar field. It is convenient however, before we proceed, to introduce a new dimensionless statefinder function in order to replace Hubble and afterwards solve numerically the system of differential equations. Before we perform such replacement it is worth rewriting the equations of motion in order to give a natural meaning in all extra terms arising, especially in the Friedmann equation (\ref{motion1}). 

In the first two equations (\ref{motion1})-(\ref{motion2}), the matter density and pressure are shown. In the present paper, we shall assume that matter is comprised of relativistic and non relativistic matter, denoted as $\rho_r$ and $\rho_{nr}$ respectively, therefore $\rho_m=\rho_r+\rho_{nr}$. Both contributions to $\rho_m$ are assumed to represent perfect fluids, meaning that the continuity equation for each component reads

\begin{equation}
\centering
\label{continuity}
\dot\rho_i+3H\rho_i(1+\omega_i)=0\, ,
\end{equation}
where $\omega_i=\frac{P_i}{\rho_i}$ is the equation of state of the respective fluid and index $i$ runs takes the values $r,nr$. Hence, matter density is written explicitly in terms of redshift as

\begin{equation}
\centering
\label{rhom}
\rho_m=\rho_{nr}^{(0)}(1+z)^3\left(1+\chi(1+z)\right)\, ,
\end{equation}
where $\rho_{nr}^{(0)}$ signifies the current density of non relativistic matter, $\rho_r^{(0)}$ the current density of relativistic matter and $\chi$ is the ratio of the aforementioned values, i.e $\chi=\frac{\rho_r^{(0)}}{\rho_{nr}^{(0)}}$. Similarly, we define as dark energy all the geometric terms derived from the elements of the gravitational action and contributing to equations (\ref{motion1}) and (\ref{motion2}). In other words we suggest that

\begin{align}
\centering
\label{rhoDE}
\rho_{DE}&=\frac{1}{2}\dot\phi^2+V+\frac{Rf_R-f}{2\kappa^2}-\frac{3H\dot f_R}{\kappa^2}+24\dot\xi H^3-9c\xi H^2\dot\phi^2\nonumber\\
&+\frac{3H^2}{\kappa^2}(1-f_R)\, ,
\end{align}
and
\begin{align}
\centering
\label{PDE}
P_{DE}&=\dot\phi^2+\frac{\ddot f_R-H\dot f_R}{\kappa^2}-8H^2(\ddot\xi-H\dot\xi)-16\dot\xi H\dot H\nonumber\\
&+2c\dot\phi\left(\xi(\dot H-3H^2)\dot\phi+2H\xi\ddot\phi+H\dot\xi H\dot\phi\right)\\\nonumber
&-\frac{2\dot H}{\kappa^2}(1-f_R)-\rho_{DE}\, ,
\end{align}
Consequently, the first equations of motion are written exactly as Friedman's equations

\begin{equation}
\centering
\label{motion4}
3H^2=\kappa^2\left(\rho_m+\rho_{DE}\right)\, ,
\end{equation}

\begin{equation}
\centering
\label{motion5}
-2\dot H=\kappa^2\left(\rho_m+P_m+\rho_{DE}+P_{DE}\right)\, ,
\end{equation}
As a result, one does not need to assume arbitrarily the existence of a new perfect fluid in order to interpret the apparent late-time acceleration
which is currently experienced, but rather dark energy is effectively the contribution of well motivated terms participating in the initial gravitational action. Now dark energy density and pressure are specified in this particular way in order to rewrite the equations of motion in the usual form. In particular $\rho_{DE}$ is used in order to replace the cosmological constant $\Lambda$ in the $\Lambda$CDM model and shall be used in order to examine whether it can produce similar if not identical results as the first. The label dark energy may seem bizarre for the time being since in order to be convinced, the equation of state, if a barotropic relation is used, must be examined. This analysis is indeed performed subsequently for some toy models and it becomes abundantly clear that the EoS is quite close to -1 thus mimicking the $\Lambda$CDM results and justifying the name dark energy. We elaborate on the EoS and the overall statefinders in the following examples. In addition, dark energy also behaves as a perfect fluid since

\begin{equation}
\centering
\label{continuityDE}
\dot\rho_{DE}+3H(\rho_{DE}+P_{DE})=0\, ,
\end{equation}
With these relations at hand, we define our new statefinder variable as

\begin{equation}
\centering
\label{yH}
y_H(z)=\frac{\rho_{DE}}{\rho_{nr}^{(0)}}\, ,
\end{equation}
and since Eq. (\ref{motion4}) was derived, it can easily be inferred that $y_H$ is connected to Hubble as

\begin{equation}
\centering
\label{yH1}
y_H=\frac{H^2}{m_s^2}-\frac{\rho_m}{\rho_{nr}^{(0)}}\, ,
\end{equation}
where $m_s^2=\frac{\kappa^2\rho_{nr}^{(0)}}{3}$ for convenience. Its numerical value which shall be used subsequently reads $m_s^2=1.87101\cdot10^{-67}$. Hence, Hubble's parameter along with its first two derivatives can be replaced by $y_H$ as shown below

\begin{equation}
\centering
\label{H1}
H^2=m_s^2\left(y_H+\frac{\rho_m}{\rho_{nr}^{(0)}}\right)\, ,
\end{equation}

\begin{equation}
\centering
\label{H'}
HH'=\frac{m_s^2}{2}\left(y_H'+\frac{\rho_m'}{\rho_{nr}^{(0)}}\right)\, ,
\end{equation}

\begin{equation}
\centering
\label{H''}
H'^2+HH''=\frac{m_s^2}{2}\left(y_H''+\frac{\rho_m''}{\rho_{nr}^{(0)}}\right)\, ,
\end{equation}
In order to examine the validity of a particular model relative to $\Lambda$CDM extra statefinder functions shall be utilized. In particular, we define the deceleration parameter $q$, the jerk $j$, the snap $s$ and $Om$ as 

\begin{align}
\centering
\label{statefinders}
q&=-1-\frac{\dot H}{H^2}&j&=\frac{\ddot H}{H^3}-3q-2\nonumber\\
s&=\frac{j-1}{3\left(q-\frac{1}{2}\right)}&Om&=\frac{\left(\frac{H}{H_0}\right)^2-1}{(1+z)^3-1}\, ,
\end{align}
As the name suggests, the statefinder parameters are functions of redshift and essentially we have

\begin{equation}
\centering
\label{q}
q(z)=-1+\frac{(1+z)H'}{H}\, ,
\end{equation}

\begin{equation}
\centering
\label{j}
j=(1+z)^2\left(\left(\frac{H'}{H}\right)^2+\frac{H''}{H}-\frac{2H'}{(1+z)H}\right)+1\, ,
\end{equation}

Finally in order to correctly interpret the extra geometric terms in equations (\ref{rhoDE}) and (\ref{PDE}) truly as dark energy density and pressure respectively, one must ascertain whether the equation of state parameter $\omega_{DE}$ (EoS) specified as the ratio of pressure over density, meaning $\omega_{DE}=\frac{P_{DE}}{\rho_{DE}}$ and the density parameter $\Omega_{DE}$ coincide with current observations. For the sake of completeness, such statefinder parameters are essentially written as functions depending solely on auxiliary function $y_H$ as shown below

\begin{equation}
\centering
\label{omega}
\omega_{DE}=-1+\frac{1+z}{3}\frac{d\ln y_H}{dz}\, ,
\end{equation} 

\begin{equation}
\centering
\label{Omega}
\Omega_{DE}=\frac{y_H}{y_H+\tilde\rho}\, ,
\end{equation}
where the tilde stands for $\tilde\rho=\frac{\rho}{\rho_{nr}^{(0)}}$. By studying the behavior of all the previous statefinders the validity of any scalar-tensor model can be inferred by direct comparison of the results with the $\Lambda$CDM model. In the following section we specify arbitrary functions in Eq. (\ref{action}) and numerically solve the system of differential equations (\ref{motion1}) and (\ref{motion3}) with respect to $y_H(z)$ and $\phi(z)$.

\section{Numerical Analysis And Validity Of The Model}
Let us now proceed with the numerical analysis of a given scalar-tensor model. Suppose that the arbitrary $f(R,\phi)$ function is minimally coupled to the scalar field and given by the expression

\begin{equation}
\centering
\label{f(R)}
f(R)=R e^{\frac{R_0}{R}}\, ,
\end{equation}
where for consistency $R_0$ is an auxiliary parameter with mass dimensions $ev^{2}$. This function is chosen since in the limit of $R\to\infty$ its contribution is negligible and thus only the scalar field drives the inflation. Concerning the scalar functions of the model we assume that they are trivially described by exponential functions of the following forms

\begin{figure}[h!]
\centering
\label{plot1}
\includegraphics[width=17pc]{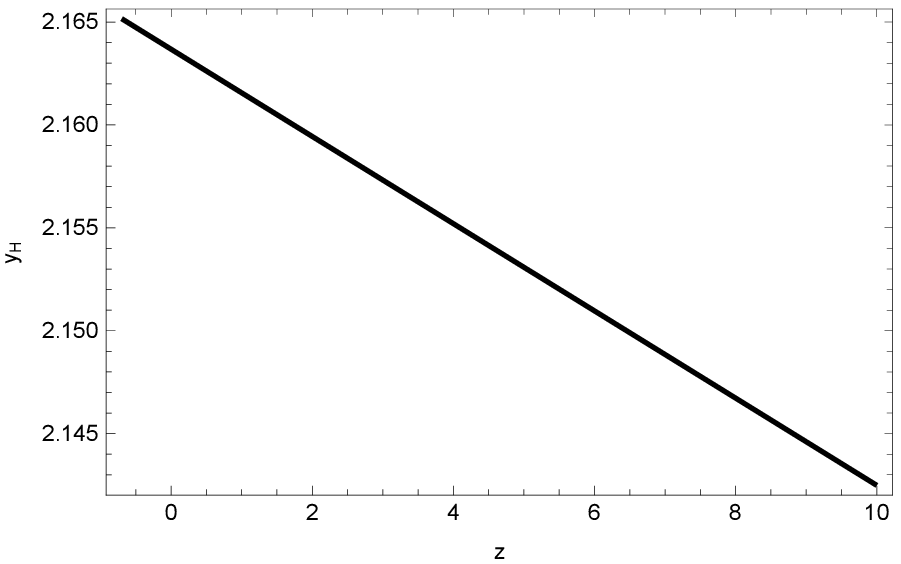}
\includegraphics[width=18pc]{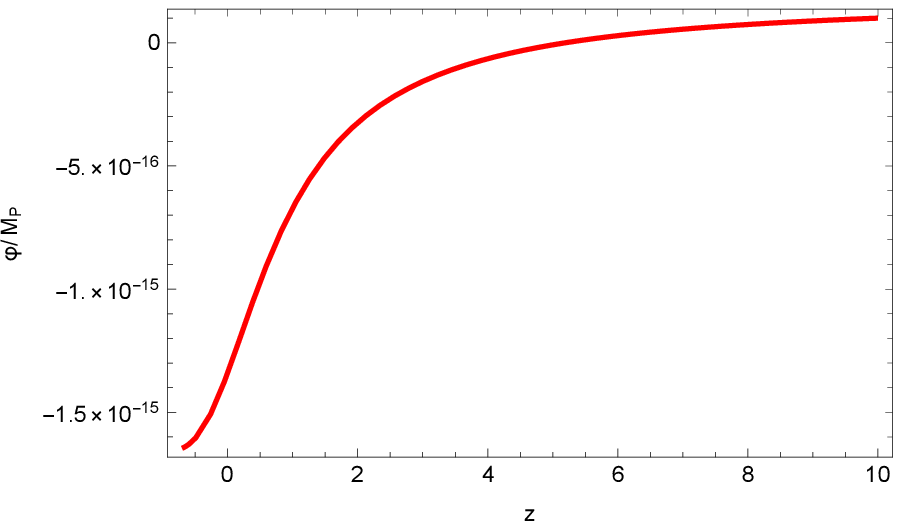}
\caption{Numerical solutions for statefinder $y_H$ (black) and the scalar field (red) normalised with respect to the Planck mass in the area [-0.9,10]. As shown the statefinder $y_H$ has a really slow evolution with respect to redshift while the scalar field transitions from positive to negative values in the late-time era.}
\end{figure}

\begin{equation}
\centering
\label{V}
V(\phi)=\Lambda^2(1-e^{\frac{\phi}{\phi_0}})\, ,
\end{equation}

\begin{equation}
\centering
\label{xi}
\xi(\phi)=\e^{-\frac{\phi}{\phi_0}}\, ,
\end{equation}  
respectively with $\Lambda$ being the cosmological constant and $\phi_0$ also being a free parameter with mass dimensions $ev$. Hereafter we substitute the values $R_0=10^{-48}ev^2$,  $\Lambda=11.895\cdot10^{-67}ev^2$ and $\phi_0=10^{-7}M_P$ with the Planck mass being $M_P=1.2\cdot 10^{28}ev$. Furthermore, the auxiliary parameter $c$ which appears in the nontrivial kinetic coupling between scalar field and curvature is assumed to be $c=\frac{1}{\Lambda}$ for appropriate units.

\begin{figure}[h!]
\centering
\label{plot2}
\includegraphics[width=17pc]{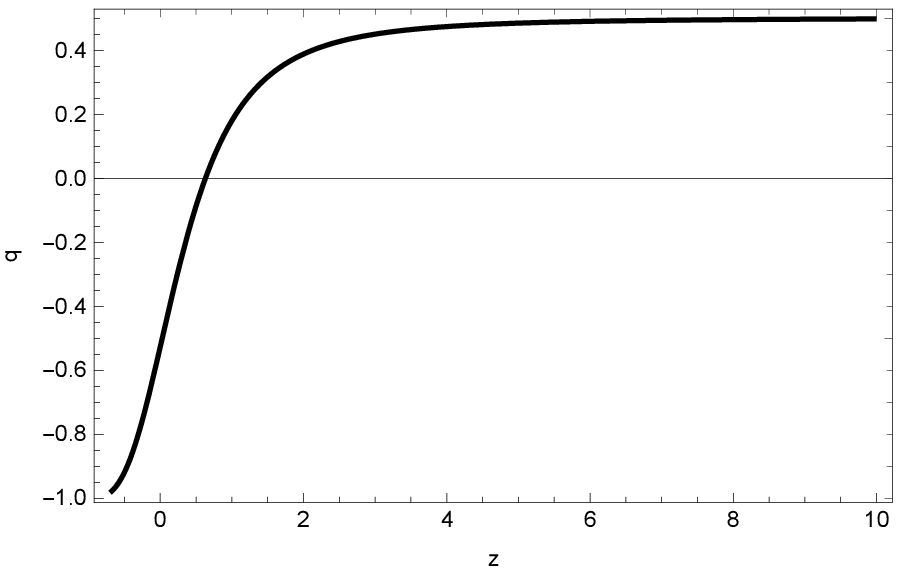}
\includegraphics[width=18pc]{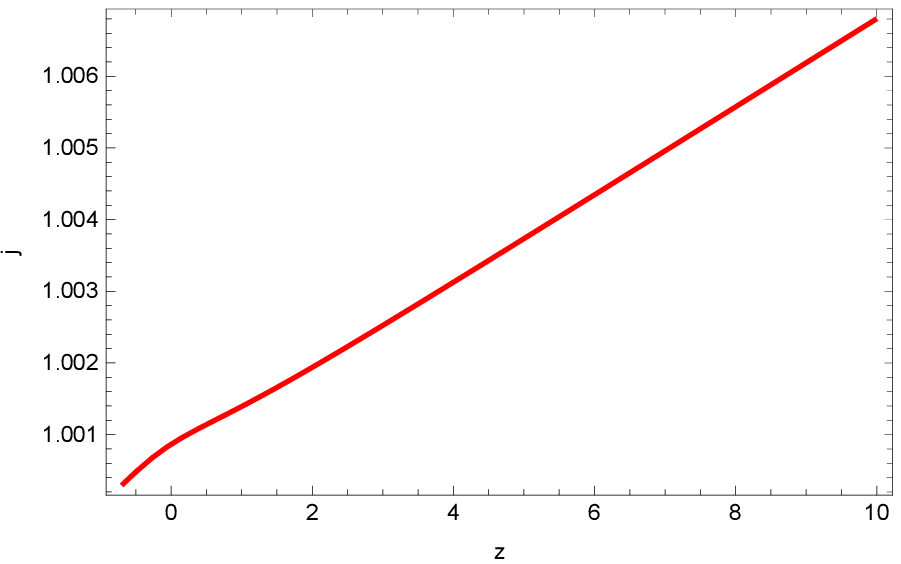}
\includegraphics[width=17pc]{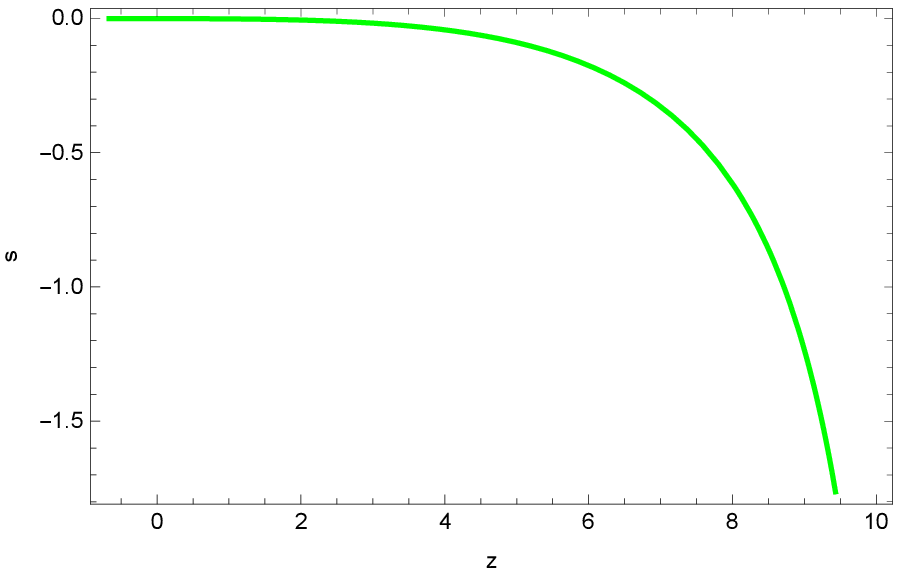}
\includegraphics[width=18pc]{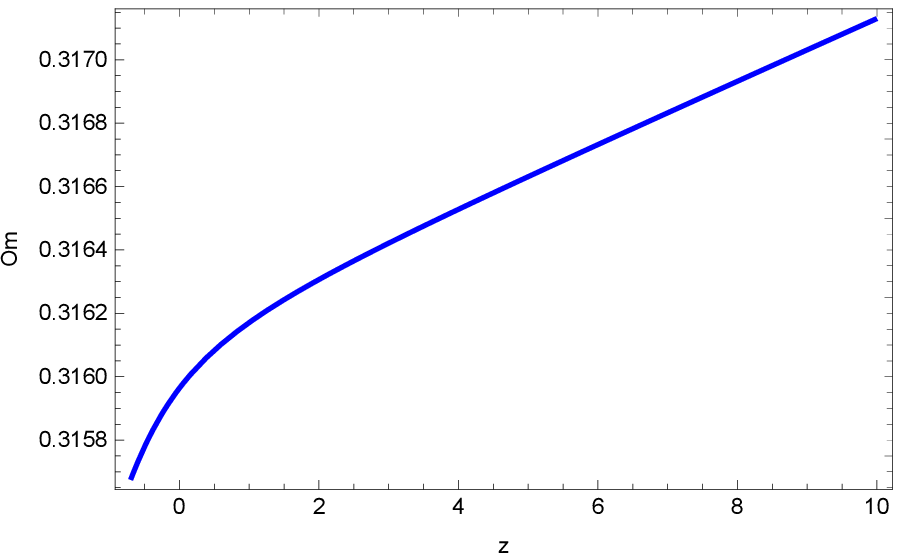}
\caption{Behaviour of the 4 aforementioned statefinder parameters in the late-time era corresponding to the area of [-0.9,10] for redshift. As shown, compatibility with the $\Lambda$CDM model is achieved for this particular set of functions and parameters, namely the deceleration parameter (black), the jerk(red), the snap (green) and $Om$ (blue).}
\end{figure}
Additionally, the initial conditions which shall be used in order to solve numerically the equations of motion (\ref{motion1}) and (\ref{motion3}) are taken to be equal to $y_H(z=10)=\frac{\Lambda}{3m_s^2}\left(1+\frac{11}{1000}\right)$ and $y_H'(z=10)=\frac{\Lambda}{3000m_s^2}$ while for the scalar field we assume that the initial conditions are $\phi(z=10)=10^{-16}M_P$ and $\phi'(z=10)=10^{-17}M_P$. Here we also need to elaborate further on the choice of initial conditions and in particular the dimensions of such choice. Given that $\dot\phi=-H(1+z)\phi'$ and $[\dot\phi]=ev^2$ while simultaneously $[H]=ev$, the initial conditions for the derivative of the scalar field must also be in units of $ev$. Therefore the Planck mass is used for both the numerical value and the derivative of the scalar field. Note that such values are not arbitrary but are taken from Ref\cite{Odintsov:2020qyw} where a different late-time model was studied and are used in order to fine tune in an essence the results and replicate the $\Lambda$CDM. Therefore, upon solving numerically equations (\ref{motion1}) and (\ref{motion3}), the functions plotted in Fig.1 emerge. As shown statefinder $y_H$ decreases with redshift whereas the scalar field increases.

In order to ascertain the validity of the model at hand a comparison between the numerical expectations of the model at hand for various statefinder parameters and the $\Lambda$CDM model \cite{Aghanim:2018eyx} is performed. We limit our work to only the 4 statefinders introduced in Eq.(\ref{statefinders}) and the dark energy parameters. The numerical analysis suggests that the current value of the deceleration parameter is $q(0)=-0.526$ which is in a relatively good agreement with the $\Lambda$CDM value shown in Table I. Moreover the jerk seems to be infinitesimally close to unity through the whole area of $[-0.9,10]$. Furthermore the snap parameter seems to be in agreement as well as its value is close to zero and last but not least the $Om$ statefinder is quite close to the current value of the matter density parameter $\Omega_m$ as it should. Finally for the dark energy parameters one observes that the EoS is infinitesimally close to $-1$ such that the relation $P_{DE}=-\rho_{DE}$ is numerically satisfied for small values of redshift  whereas the current dark energy density parameter taken from Table I is expected to have an acceptable with the observations value given that essentially $\Omega_{DE}(0)+\Omega_{m}(0)\simeq1$, the Friedmann constrained is indeed fulfilled. Thus the current model is in a good agreement with the $\Lambda$CDM model and could be a valid candidate for the description of the late-time acceleration. The aforementioned statefinders referring to the evolution of the universe are depicted in Fig. 2 whereas the dark energy parameters are shown in Fig.3 from which it can easily be inferred that compatibility with the Planck data \cite{Aghanim:2018eyx} is achieved.

\begin{figure}[h!]
\centering
\label{plot3}
\includegraphics[width=17pc]{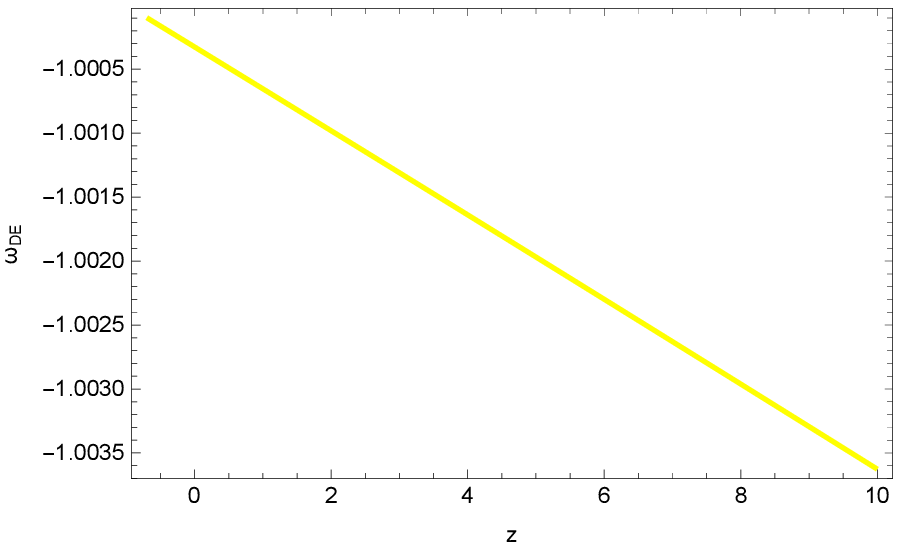}
\includegraphics[width=18pc]{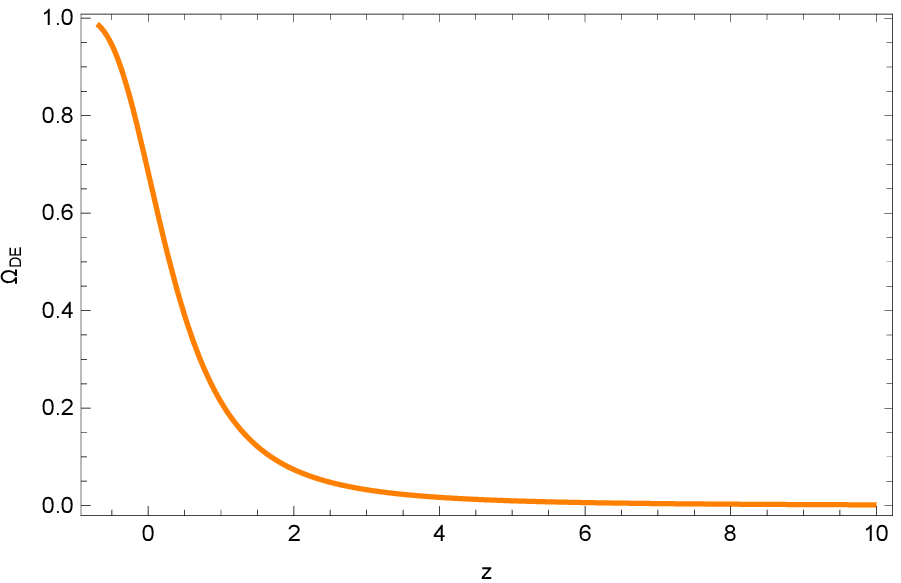}
\caption{Dark energy EoS (yellow) and density parameter (orange) as functions of redshift. As showcased, the EoS is quite close to the value of -1 and moreover the dark energy density parameter is effectively zero during the final stages of the matter dominated era and starts increasing with a fast rate up to unity in the current accelerating phase the universe experiences.}
\end{figure}

As a final note, it is worth investigating the numerical value of the gravitational wave velocity in order to ascertain whether the aforementioned models are in agreement with the GW170817 event. For such a framework, the aforementioned velocity is given by the following expression

\begin{figure}[h!]
\centering
\label{plot4}
\includegraphics[width=17pc]{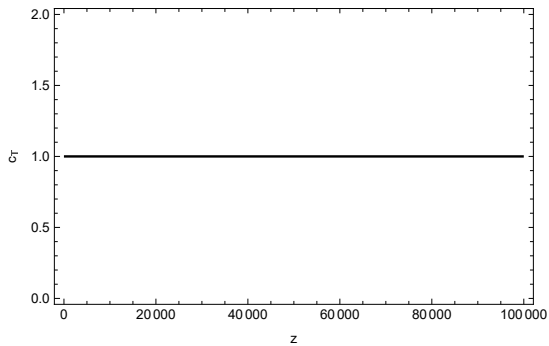}
\caption{Gravitational wave velocity as a function of redshift running from the $-0.9$ up to $z=10^5$, covering the late-time epoch, matter dominated and radiation dominated eras. For the first model the velocity is identically equal to unity therefore compatibility with the GW170817 event is achieved. Overall the model is indeed compatible with the $\Lambda$CDM.}
\end{figure}

\begin{equation}
\centering
\label{cT}
c_T^2=1-\frac{Q_2}{Q_1}\, ,
\end{equation}
where $Q_2=8(\ddot\xi-H\dot\xi)+c\xi\dot\phi^2$ and $Q_1=f_R-8\dot\xi H+c\xi\dot\phi^2$. If one examines the previous 2 models then it becomes apparent that the velocity in both cases is infinitesimally close to unity in the whole regime of $[-0.9,10]$ for redshift therefore compatibility with the GW170817 event is indeed guaranteed. In Fig.4 one can indeed see that the velocity of gravitational waves remains unchanged throughout three seemingly different cosmological eras. As a result, gravitons remain massless from the late-time era to the radiation dominated era for the first model. This however is not guaranteed for every model, as shown in the following example.

\begin{table}
\centering
\begin{tabular}{|c|c|c|}\hline
Statefinders&Numerical Results& $\Lambda$CDM Value\\ \hline
q(0)&-0.52605&-0.535\\ \hline
j(0)&1.00087&1\\ \hline
s(0)&-0.00028&0\\ \hline
Om(0)&0.31596&0.3153$\pm$0.007\\ \hline
$\omega_{DE}(0)$&-1.00033&-1.018$\pm$0.31\\ \hline
$\Omega_{DE}(0)$&0.683845&0.6847$\pm0.0073$\\ \hline
\end{tabular}
\caption{Comparison between the numerical and the observed value of the kinetic-coupled $f(R)=R e^{\frac{R_0}{R}}$ model and the $\Lambda$CDM. As shown the numerical results are in a good agreement with observations.}
\end{table}

Suppose now a different scenario in which the auxiliary functions of the model are specified as follows

\begin{figure}[h!]
\centering
\label{plot5}
\includegraphics[width=17pc]{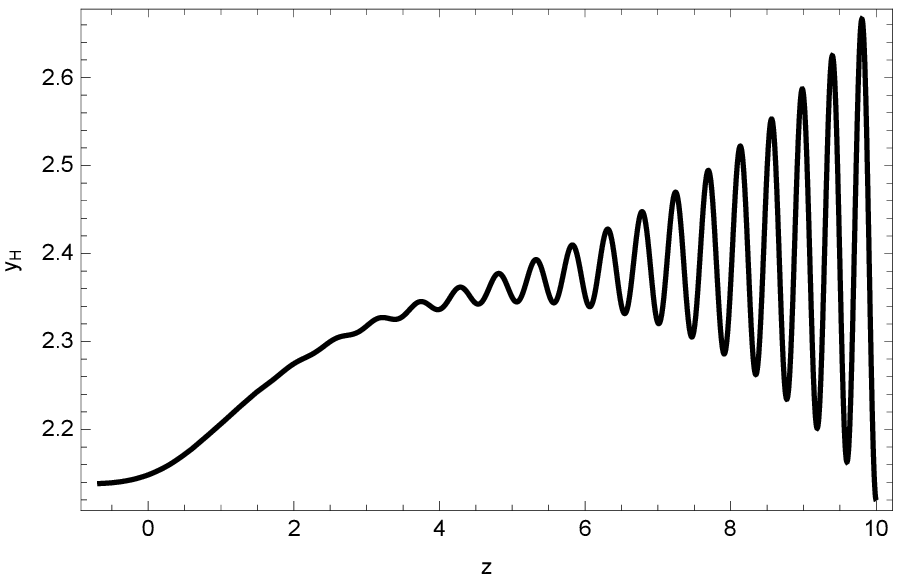}
\includegraphics[width=18pc]{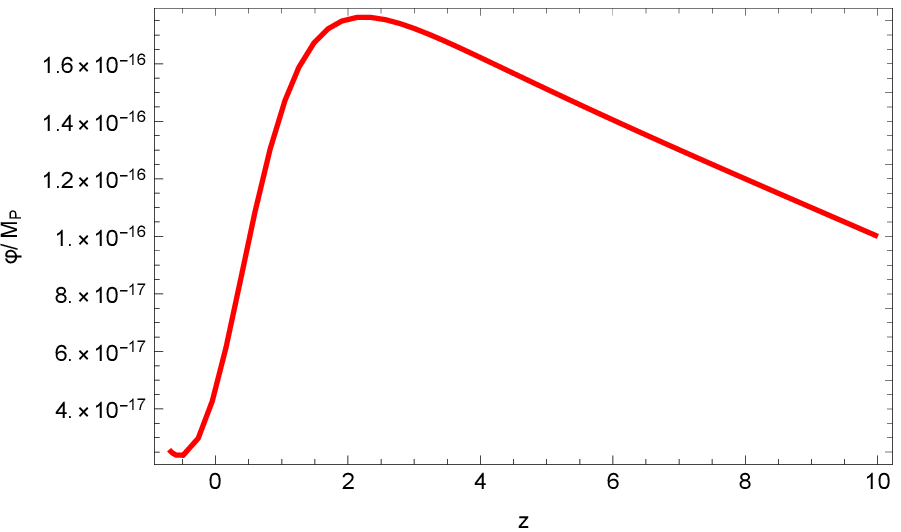}
\caption{Numerical solutions for statefinder $y_H$ and the scalar field in the area [-0.9,10]}
\end{figure}

\begin{equation}
\centering
\label{f2}
f(R,\phi)=\left(1+\left(\frac{\phi}{10M_P}\right)^2\right) R-2\Lambda \left(\frac{R}{\Lambda}\right)^{\frac{1}{100}}\, ,
\end{equation}

\begin{figure}[h!]
\centering
\label{plot6}
\includegraphics[width=17pc]{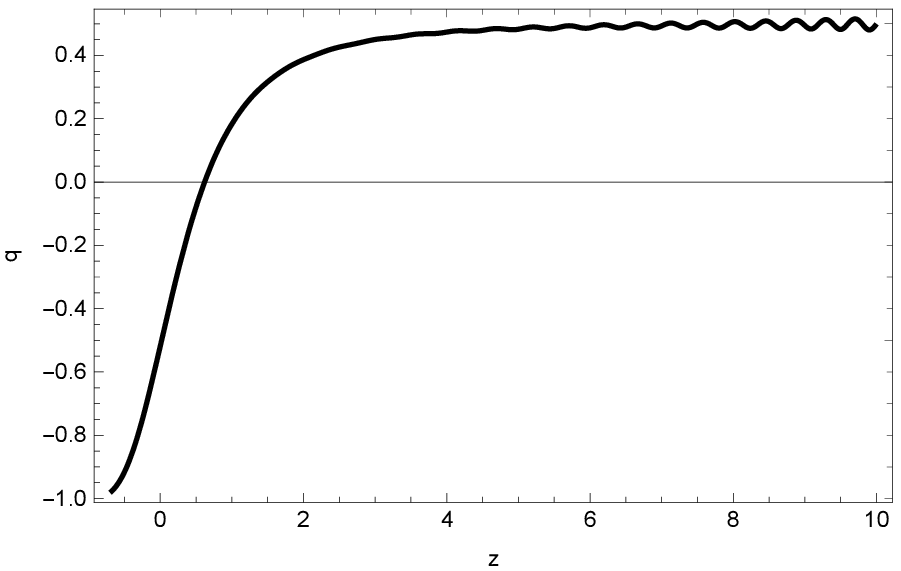}
\includegraphics[width=18pc]{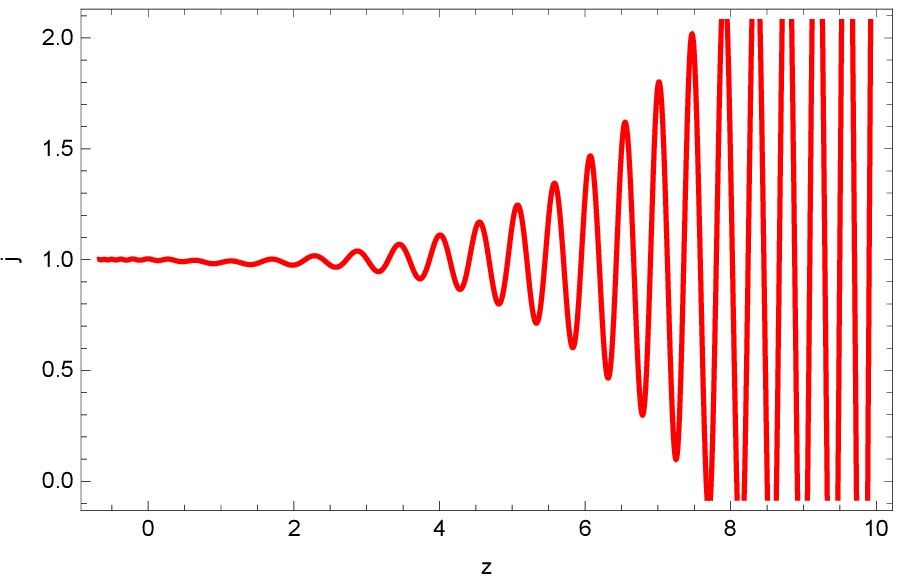}
\includegraphics[width=17pc]{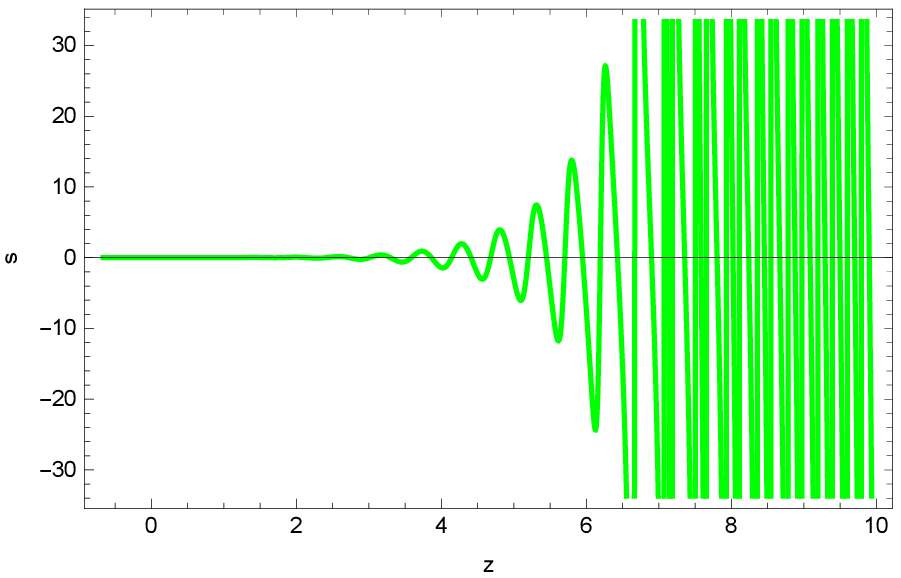}
\includegraphics[width=18pc]{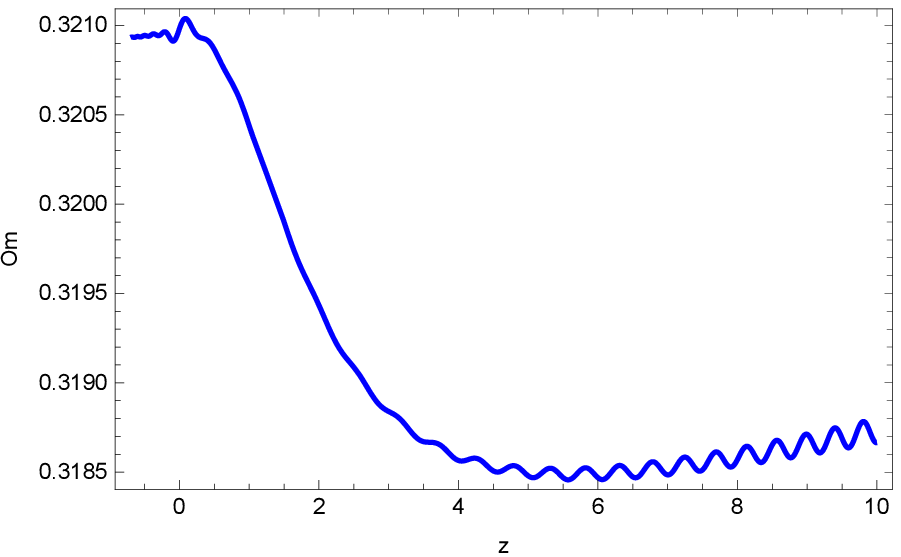}
\caption{Deceleration (black), jerk (red), snap (green) and $Om$ (blue) as functions of redshift in the area of [-0.9,10]. As shown the statefinders carry also dark energy oscillations with higher derivative term being affected greater that the rest parameters.}
\end{figure}

\begin{equation}
\centering
\label{V2}
V(\phi)=(10\Lambda)^2\left(\frac{\phi}{10^{-38}M_P}\right)^4\, ,
\end{equation}
with $\xi(\phi)$ having the exact same exponential form as in Eq.(\ref{xi}) and now $c=\frac{10^{90}}{MP^2}$. The $f(R)$ model is taken from Ref \cite{Odintsov:2020nwm} and is known for producing dark energy oscillations throughout early stages of the accelerating era. The exponent is strictly lesser than unity therefore its contribution is important in the limit of $R\to0$ while during the inflationary era the non minimal part shall be dominant. The aforementioned dark energy oscillations rise with the increase of redshift so essentially the matter dominated era is infested with such oscillations. Moreover, several inclusions cannot nullify them, see for instance Ref \cite{Odintsov:2020nwm,Odintsov:2021nim}. Hence it is interesting to examine the impact of higher order curvature corrections non minimally coupled to the scalar field on the overall phenomenology assuming now that $y_H(z=10)=\frac{\Lambda}{3m_s^2}$, $y_H'(z=10)=-\frac{y_H(z=10)}{1000}$, $\phi(z=10)=10^{-16}M_P$ and $\phi'(z=10)=-10^{-17}M_P$ one finds that the previous model is indeed compatible with the $\Lambda$CDM model however the dark energy oscillations are not cancelled by the inclusion of such terms. Numerical solutions of statefinder $y_H$ and the scalar field $\phi$ are shown in Fig.5 from which it becomes abundantly clear that dark energy oscillations are indeed present. All the current numerical values of the statefinders for the second model are depicted in Table.II from which it can easily be ascertained that compatibility with the Planck data \cite{Aghanim:2018eyx} is indeed achieved but at a cost of dark energy oscillations described by an increasing with redshift amplitude. Note also that higher derivatives of $y_H$ are affected even more as shown in Fig. (6) where the 4 statefinders are showcased while in Fig.7 the respective dark energy EoS and density parameter are depicted. As expected, higher derivatives of $y_H$ manage to enhance the aforementioned oscillations.

\begin{figure}[h!]
\centering
\label{plot7}
\includegraphics[width=17pc]{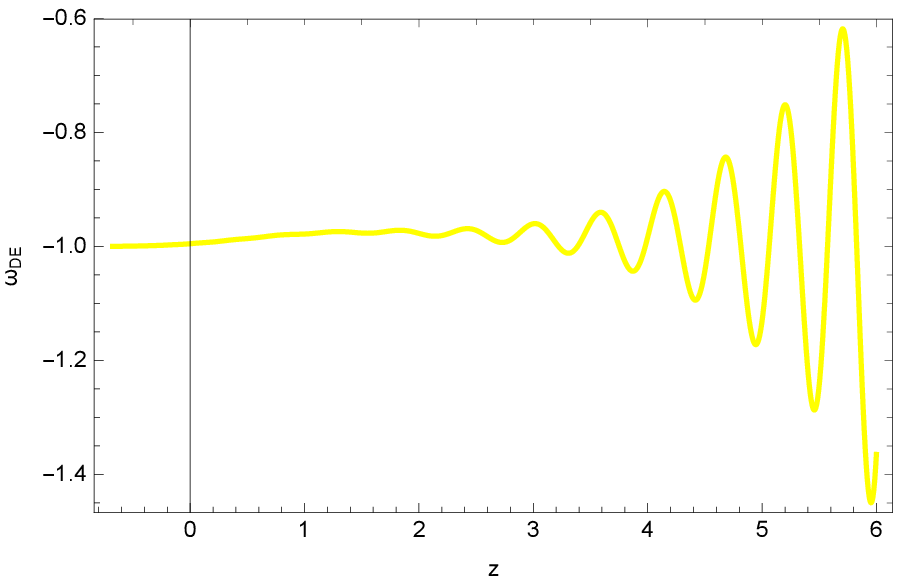}
\includegraphics[width=18pc]{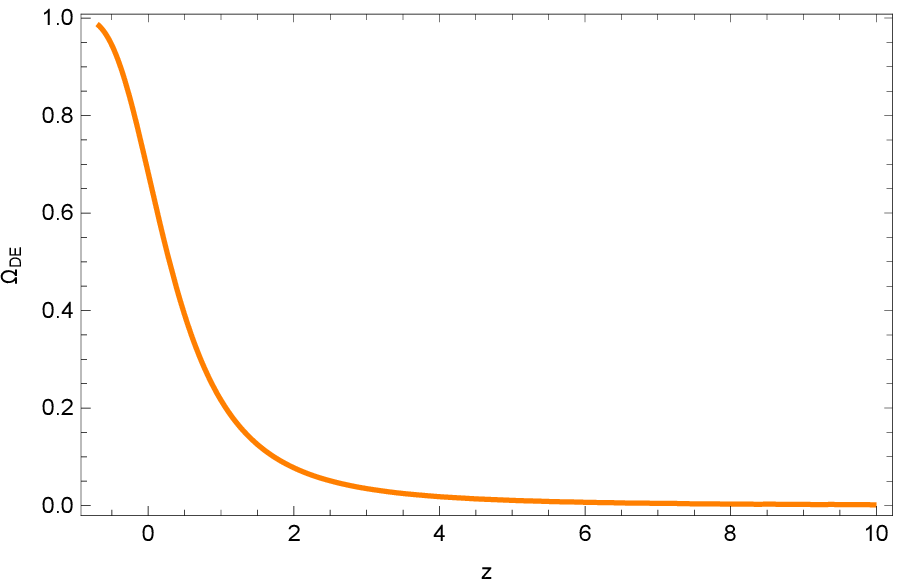}
\caption{Dark energy EoS and density parameter as functions of redshift}
\end{figure}

\begin{table}
\centering
\begin{tabular}{|c|c|c|}\hline
Statefinders&Numerical Results& $\Lambda$CDM Value\\ \hline
q(0)&-0.518535&-0.535\\ \hline
j(0)&1.00373&1\\ \hline
s(0)&-0.00122&0\\ \hline
Om(0)&0.32098&0.3153$\pm$0.007\\ \hline
$\omega_{DE}(0)$&-0.995219&-1.018$\pm$0.31\\ \hline
$\Omega_{DE}(0)$&0.682313&0.6847$\pm0.0073$\\ \hline
\end{tabular}
\caption{Numerical versus observational current values of statefinder parameters between the non-minimally coupled power-law $f(R,\phi)$ model and the $\Lambda$CDM. It becomes apparent that even though dark energy oscillations are present, compatibility is achieved.}
\end{table}

\begin{figure}[h!]
\centering
\includegraphics[width=17pc]{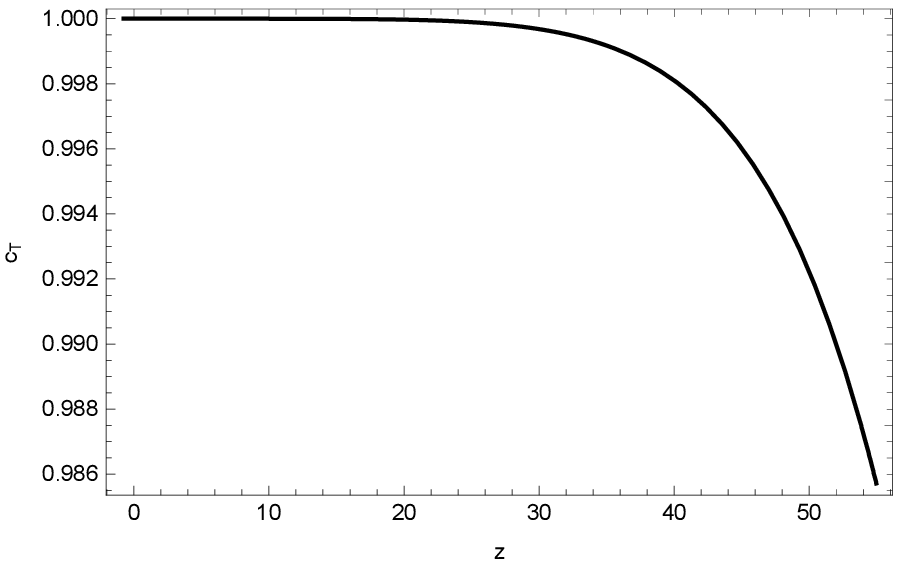}
\includegraphics[width=17pc]{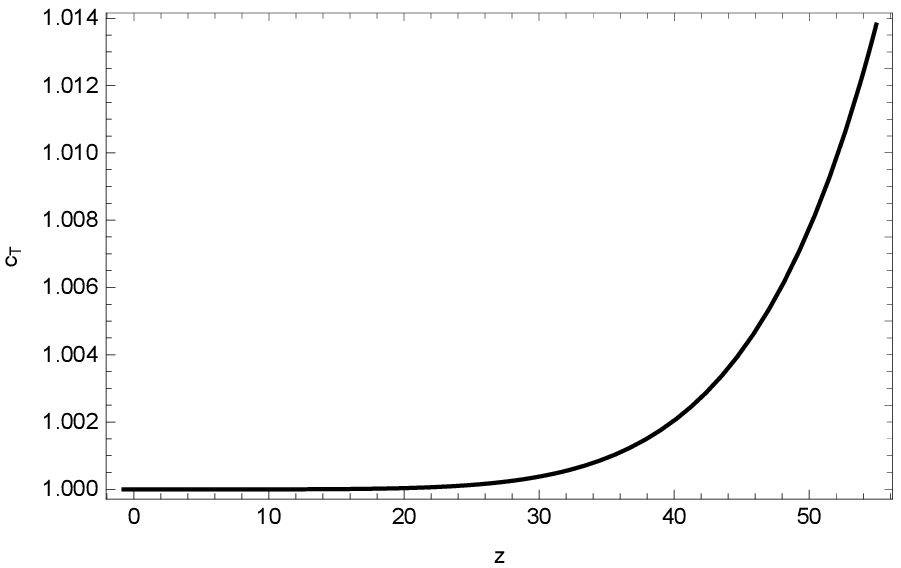}
\caption{Evolution of the gravitational wave velocity from late-time to matter dominated era for the case of $c=\frac{10^{90}}{M_P^2}$ (upper) and $c=-\frac{10^{90}}{M_P^2}$(lower). It can easily be inferred that both the sign and the value of parameter $c$ affects the overall phenomenology. Both descriptions must be rendered nonviable given that in the first compatibility with the GW170817 event is violated whereas in the latter causality is violated.}
\label{plot8}
\end{figure}

Before closing this section it is worth mentioning that the power-law model makes in fact an additional prediction for the propagation velocity of gravitational waves that renders it as a non viable model. Studying tensor perturbations suggests that $c_T$ actually decreases with redshift therefore it violates the findings of the GW170817 event as we can clearly see a deviation between $c_T$ and the velocity of light during the matter dominated era. Moreover, a negative vaue of parameter $c$ may lead to violation of causality as a negative $c$ implies a velocity faster than light. As a result the model is incompatible with observations, however one can easily modify such findings by altering the numerical value of parameter $c$. In fact, when $c$ decreases no differences between the velocity of light and that of gravitational waves is observed. Therefore, $c$ in principle should be constrained in the early era from inflationary models where $\mathcal{G}$ and $G^{\mu\nu}\nabla_\mu\phi\nabla_\nu\phi$ are more dominant than the $R^{\frac{1}{100}}$ term. A comparison with the findings of Ref \cite{Oikonomou:2020tct} suggests that for the case of $c=-\frac{1}{M_P^2}$, or at least at such order of magnitude approximately, both the inflationary era and the late-time era are described decently for the case of a $\phi^4$ scalar potential however in the previous reference a linear Gauss-Bonnet scalar coupling function and a minimally coupled gravity were implemented, where in addition a constraint on the gravitational velocity was implemented. Overall, a proper designation of auxiliary parameters is needed to account for the possibility of $c_T\neq1$. This would also imply a massive graviton during the matter dominated era which as shown from the first part Fig.\ref{plot8} the mass is increasing with redshift.

Finally, a brief explanation of why the case of positive but quite large $c$ is labelled as nonviable in Fig.(8) is needed. One could argue that the GW170817 event was observed in the accelerating epoch of the universe therefore the propagation velocity could be lesser than unity in natural units in different cosmological eras where essentially an important dynamical evolution of the scalar field could be observed. Although mathematically speaking the aforementioned statement may be sensible, from a physical point of view one may need to introduce a mechanism which gives mass to gravitons in the rest cosmological eras, both primordially in the inflationary era and in radiation/matter dominated era, but renders them massless during the accelerating phase of the universe. To our knowledge, no such mechanism exists therefore a reasonable assumption is to postulate that the gravitational waves propagate with the velocity of light in each cosmological era of interest therefore a reparametrisation of $c$ is needed in order to achieve compatibility. Indeed,using a different value for $c$, one which is lesser in absolute value than the current restores viability.

\section{Conclusions}
In the present article we studied the late-time behaviour of certain string inspired models of gravity. Initially the framework was constructed by performing essentially two changes, the first being a variable change that enables one to work with redshift instead of cosmic time t and the second is a change of functions thus instead of Hubble's parameter, a dimensionless statefinder connected with the definition of dark energy density is used. By appropriately modifying the equations of motion in order to account for such changes, numerical solutions for the late-time era were extracted for two models of interest. In both cases we find that compatibility with the $\Lambda$CDM description can easily be obtained by properly specifying not only the auxiliary functions but also the free parameters of the model. Moreover, for the case of the power-law $f(R)$, apart from the presence of persisting dark energy oscillations in higher redshifts, it becomes abundantly clear that even though in the late-time era the $f(R)$ contribution is more dominant, while the string corrective terms $\xi(\phi)\mathcal{G}$ and $\xi(\phi)g^{\mu\nu}\nabla_\mu\phi\nabla_\nu\phi$ are expected to be dominant during the inflationary era, extra caution is needed as a naive designation of certain parameters may spoil the propagation velocity of gravitational waves in the current accelerating era. To account for such possibility, a more thorough analysis of the inflationary era is needed such that both eras may be unified properly. Upon unification, an analysis of the energy spectrum of primordial gravitational waves in order to examine whether it is possible to describe a potential future signal detection in second and third order generation detectors such as LISA and NANOGrav from such scalar-tensor models. We aim to address this task in a future work.

\section*{Acknowledgments}
The author would like to express his gratitude towards Dr. V.K.Oikonomou for his various comments and suggestions on the Horndeski subclass model that elevated the status of the paper.

\end{document}